\documentclass[twocolumn]{aastex7}
\usepackage{xcolor}
\usepackage{xspace}
\usepackage{amsmath}
\usepackage{CJKutf8}
\usepackage{threeparttable}
\usepackage{tabularx}
\usepackage{makecell}
\usepackage{multirow}
\usepackage{orcidlink}
\usepackage{natbib}
\usepackage{enumerate}
\usepackage{amssymb}
\usepackage{graphicx}
\graphicspath{{Figures/}}

\begin{document}
\newcommand{\sqb}[1]{\left[#1\right]}
\newcommand{\pr}[1]{\left(#1\right)}
\newcommand{\memotl}[1]{ \textcolor{magenta}{$TL[$ \bf{#1} $]$}}
\newcommand{\memoday}[1]{ \textcolor{blue}{$DAY[$ \bf{#1} $]$}}
\newcommand{\rev}[1]{\textcolor{orange}{#1}}

\begin{CJK*}{UTF8}{gbsn}
\title{A Massive Hot-Jupiter Companion that Disfavors Giant Planet Formation Beyond the Water-Ice Line}
\author{Eritas Yang (杨晴) \orcidlink{0009-0005-2641-1531}}
\affiliation{Department of Astrophysical Sciences, Princeton University, 4 Ivy Lane, Princeton, NJ 08540, USA}
\email{eritas.yang@princeton.edu}

\author{Tiger Lu (陆均) \orcidlink{0000-0003-0834-8645}}
\altaffiliation{Flatiron Research Fellow}
\affiliation{Center for Computational Astrophysics, Flatiron Institute, 162 5th Avenue, New York, NY 10010, USA}
\email{tlu@flatironinstitute.org}

\author{Daniel A. Yahalomi (丹雅浩) \orcidlink{0000-0003-4755-584X}}
\altaffiliation{Juan Carlos Torres Postdoctoral Fellow}
\affiliation{Kavli Institute for Astrophysics and Space Research, Massachusetts Institute of Technology, Cambridge, MA 02139, USA}
\email{yahalomi@mit.edu}

\author{Joshua N. Winn (温乔书) \orcidlink{0000-0002-4265-047X}}
\affiliation{Department of Astrophysical Sciences, Princeton University, 4 Ivy Lane, Princeton, NJ 08540, USA}
\email{jnwinn@princeton.edu}

\begin{abstract}
We report evidence for a brown-dwarf companion with mass $34^{+30}_{-11}~M_{\rm J}$ in the KELT-20 system, in which an ultra-hot Jupiter transits an A2-type star. 
The companion's properties are inferred from a joint analysis of
astrometric accelerations and transit timing variations, and its present-day orbit 
imposes dynamical limits on where the hot Jupiter
could have formed. Given the star's current luminosity,
the water-ice line is expected at $\sim$8--15~au,
but the companion's inferred pericenter distance
of a few au would lead to orbit crossing
or long-term instability for any planet
formed at such distances.
If the companion formed early and remained near its current orbit over the system's lifetime, the proto--hot Jupiter must have formed within $\sim$3.7~au to avoid orbit crossing, and within $\sim$1.5~au to remain dynamically stable over the system's lifetime.
These results disfavor formation beyond the ice line
and point instead to formation at smaller
orbital distances followed by inward migration.
\end{abstract}

\section{Introduction}
The origin of hot Jupiters remains an open problem. One possibility is that they form \textit{in situ}, near their present-day orbits, either through gravitational instability -- the fragmentation of the protoplanetary disk into bound clumps \citep{boss1997GI} -- or through core accretion, in which a solid core grows and accretes a massive gaseous envelope \citep[e.g.,][]{Perri_1974, pollack1996accretion}.
Although both of these {\it in situ} formation mechanisms
are physically plausible \citep[e.g.,][]{Batygin_2016,Boley_2016} they are generally disfavored.

The problem with gravitational instability
at small orbital separations
is that the disk is expected to be too hot and diffuse
to satisfy the Toomre criterion for instability \citep{Toomre_1964}.
Even if the disk were massive enough to become gravitationally unstable, any fragments that form close to the star would likely be disrupted by shear before they could cool and contract \citep{Rafikov_2005}.
Formation via core accretion faces a different challenge:
forming the $\sim 10\,M_\oplus$ core needed for runaway gas accretion requires both a sufficient reservoir of solids and a growth time shorter than the gas-disk lifetime. At small orbital separations, these conditions are difficult to meet \citep{Lee_2016,dawson2018origin}.

The prevailing view is therefore that most hot Jupiters form beyond the water-ice line and subsequently migrate inward to their present-day orbits \citep{lin1996orbital, Bodenheimer_2000, ida2004deterministic, dawson2018origin}.
Disk-driven migration can transport a giant planet from beyond the ice line to the inner disk edge within the $\sim$Myr lifetime of the protoplanetary disk \citep[e.g.,][]{Ida_2008,Baruteau_2014,Heller_2019}. Alternatively, high-eccentricity migration can occur when dynamical interactions with exterior companions excite large eccentricities that are later damped by tides, shrinking and circularizing the orbit \citep[e.g.,][]{Wu2003,Fabrycky2007, Wu2011,Beauge2012}.

KELT-20\,b, also known as MASCARA-2\,b, is an
ultra-hot Jupiter transiting a 1.9\,$M_\odot$ A2 star
with an orbital period
$P_b=3.474$ days, semi-major axis $a_b=0.054$ au, radius
$R_b=1.7~R_{\rm J}$, and mass upper limit
$m_b<3.4~M_{\rm J}$ \citep{Lund_2017, Talens_2018}.
Its sky-projected obliquity is $\lambda = 3.4 \pm 2.1^\circ$, suggesting that the planetary orbit is closely aligned with the stellar spin axis \citep{Lund_2017}.
With an age of $58 \pm 5$ Myr \citep{Distler_2026}, KELT-20 is among the youngest known hot Jupiter systems, offering an unusually early view of the processes that shape the orbits of close-in giant planets. 
The host star shows a significant astrometric acceleration, suggesting the presence of an additional companion.

In this work, we use long-term trends in astrometric and transit timing data to confirm the presence of the distant companion and estimate its properties. Our fits favor a brown dwarf on a ${\sim}$few au orbit. We discuss some implications for the planet's formation and migration history given the presence of this companion.

\section{Data}
\subsection{Astrometric acceleration}
We adopted absolute astrometry for KELT-20 from the Hipparcos-Gaia Catalog of Accelerations \citep{Brandt_2021}, which places
measurements from Hipparcos \citep{perryman1997hipparcos} and Gaia \citep{gaia2016} onto a common reference frame. The catalog
provides three independent estimates of the proper motion: one measured by Hipparcos, one by Gaia, and a long-term average proper motion obtained from the displacement vector that joins the Hipparcos and Gaia mean positions over their $\sim$25-year baseline.

\begin{figure}[t]
    \centering
    \includegraphics[width=1.0\linewidth]{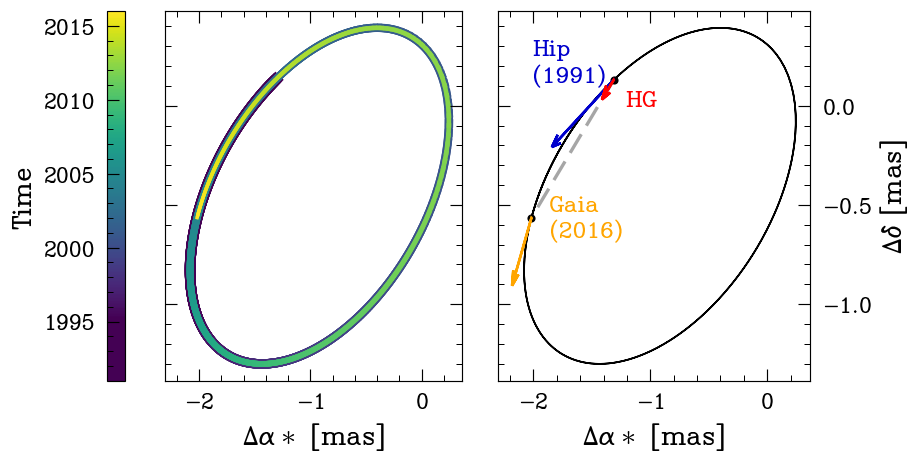}
    \caption{Possible sky-projected orbit of KELT-20 induced by a companion, shown for the best-fit solution
    (Section~\ref{ssec:joint}).
    \textbf{Left:} Motion of the host star between the Hipparcos and Gaia epochs (1991--2016), color-coded by time. The axes show right ascension (RA; $\alpha*\equiv \alpha\cos{\delta}$) and declination (Dec; $\delta$), with the origin at the center of mass.
    \textbf{Right:} Illustration of proper-motion
    vectors from Hipparcos (blue), Gaia (orange), and Hipparcos–Gaia (HG; red). 
    The HG vector points along the displacement vector between Hipparcos and Gaia epochs (gray dashed line). It is much shorter
    than the others because it is
    based on the net displacement over $\sim$25 years,
    during which the star completed several orbits
    and returned to nearly its original position as shown in the left panel.}
    \label{fig:HGCA_schematic}
\end{figure}

KELT-20 exhibits a $7\sigma$ discrepancy ($\chi^2 = 62$ for 2 degrees of freedom) between the Gaia proper motion and the long-term average proper motion, indicating a significant sky-projected acceleration of the star (see Figure~\ref{fig:HGCA_schematic} for a schematic). We interpret this acceleration as evidence for an unseen companion.

\subsection{Transit timing variations \label{ssec:TTV}}

In the presence of a long-period massive companion, the center of mass
of the star and hot Jupiter accelerates, leading to a gradual change
in the observed interval between successive transits due to the
changing light-travel-time (the R{\o}mer effect).
We analyzed the transit timing variations (TTVs) to seek such a
signal and to further constrain the properties of the unseen
companion.

We used observations from the Transiting Exoplanet Survey Satellite (TESS; \citealt{TESS}), which captured 44 transits of KELT-20\,b between 2019 and 2024.
We retrieved the SPOC PDCSAP TESS light curves from MAST
using \texttt{lightkurve}\footnote{ \url{https://github.com/lightkurve/lightkurve}} \citep{Lightkurve} and processed them with the code presented by \citet{Ivshina2022}.
In brief, for each sector, we retained only the data within $1.5$ transit durations of each expected transit midpoint. We discarded transits with insufficient in-transit or out-of-transit coverage. Each remaining transit was locally detrended by fitting the out-of-transit data with a polynomial of degree 1--3, selected by the Bayesian Information Criterion. We then fit the phase-folded light curve with the analytic transit model of \citet{Mandel2002} to determine the transit-shape parameters for each sector. Individual transits were subsequently fit with these shape parameters fixed, allowing only the mid-transit time and the parameters of a linear detrending function to vary.
Transits for which the best-fit model had an excessively high $\chi^2$ ($p<0.01$) were disregarded.

\begin{figure}[b]
    \centering
    \includegraphics[width=0.9\linewidth]{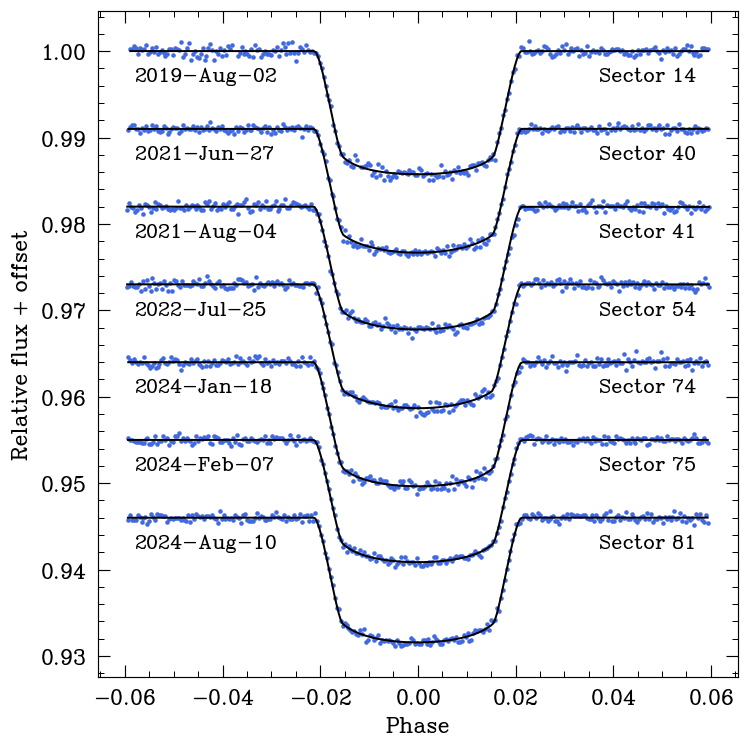}
    \caption{Individual detrended transit light curves
    of KELT-20\,b observed by TESS. Each row shows a representative detrended light curve from one sector. The blue points are TESS data and the black curves are best-fit models.}
    \label{fig:TESS_TTV}
\end{figure}

\begin{figure*}
    \centering
    \includegraphics[width=0.74\linewidth]{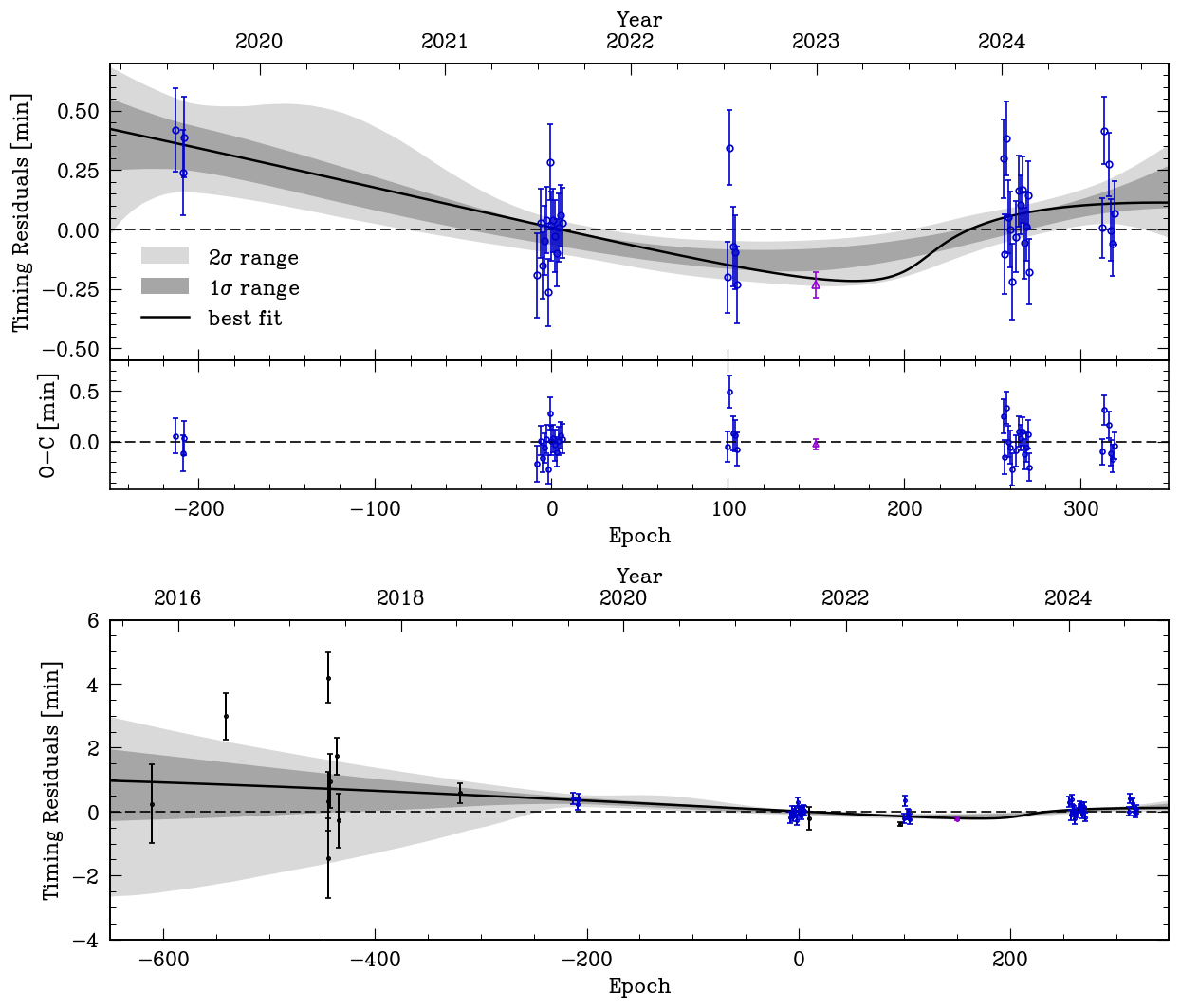}
    \caption{Transit timing variations of KELT-20\,b relative to a constant-period ephemeris. The period and mid-transit time at the zero epoch are listed in Table~\ref{tab:prior_posterior}.
    \textbf{Top:} Data used in the joint fit. Blue points are TESS measurements, and the purple point is the HST/UVIS measurement from \citet{Chachan_2025}.
    The solid black curve shows the best-fitting companion model ($\chi^2 = 48$ for 37 degrees of freedom), and the shaded regions show the $1\sigma$ and $2\sigma$ predictive intervals.
    The lower subpanel shows the residuals after subtracting the best-fitting model.
    \textbf{Bottom:} Same model shown over the longer time baseline, including earlier literature mid-transit times that were excluded from the fit because of possible lower accuracy.
    These excluded measurements are shown as black points and listed in Table~\ref{tab:past_data}.
    Most of the earlier measurements are consistent with the model despite not having been included in the fit.}
    \label{fig:TTV_residuals}
\end{figure*}

Representative detrended light curves from each sector are shown in Figure~\ref{fig:TESS_TTV}. 
We also included the mid-transit time reported by \citet{Chachan_2025}, obtained with HST/UVIS on 2022 November 23.
As a null test, we fit the combined transit times with a constant-period model.
The best-fitting solution gives $\chi^2 = 82$ for 41 degrees of freedom,
corresponding to $p = 10^{-4}$, a poor fit. 
The residuals (Figure~\ref{fig:TTV_residuals}, top panel)
show departures from a constant period that are qualitatively consistent with
a changing light-travel time caused by the line-of-sight acceleration of the
host star. A leave-one-out test shows that the HST/UVIS timing has substantial
leverage: when this point is excluded, the evidence for departures from a
linear ephemeris is weakened to $p = 0.02$. We therefore interpret the current TTV signal cautiously and emphasize its dependence on this high-precision timing
measurement.

Additional mid-transit times dating back to 2015 have been reported in the literature. 
Out of concern that these early measurements are generally less reliable, we excluded them 
from the primary analysis. However, they are listed in Appendix~\ref{app:sup} and 
shown in the bottom panel of Figure~\ref{fig:TTV_residuals}.

\subsection{Joint analysis \label{ssec:joint}}

We performed a joint fit to the astrometric and TTV data. For both Hipparcos and Gaia DR3, the cataloged proper motions are the slopes of a best-fit linear trajectory over the mission duration, rather than the instantaneous velocity at a single epoch. Accordingly, we computed the modeled Hipparcos and Gaia proper motions by fitting a linear trajectory
to a set of model-predicted sky positions, uniformly sampled in time over each mission window.

Posterior distributions for the model parameters were inferred using dynamic nested sampling \citep{Higson_2019}, as implemented in the \texttt{dynesty}\footnote{\url{https://github.com/joshspeagle/dynesty}} package \citep{dynesty}. The adopted priors are listed in Table~\ref{tab:prior_posterior}. To ensure convergence and robustness, we performed 15 independent runs, each targeting an effective posterior sample size of $N_{\rm eff}=10^5$.

\begin{table*}[]
    \centering
    \caption{Posterior constraints for the orbital and astrometric parameters of the KELT-20 system. Parameters associated with the companion are denoted by the subscript ``1''.}
    \begin{tabular}{l l r r}
    \hline
    Parameter & Description & Prior & Posterior\\
    \hline
    $t_0$ (BJD$-2457000$) & mid-transit time at reference epoch & $\mathcal{U}(2421, 2427)$ & $2424.297687^{+0.000018}_{-0.000019}$ \\
    $P_b$ (days) & period of the planet & $\mathcal{U}(3.4740,3.4742)$ & $3.4741013^{+0.0000011}_{-0.0000007}$\\
    $\frac{m_1}{a_1^2}$ $\pr{\rm \frac{M_J}{au^2}}$ & astrometric acceleration measure & $\log\mathcal{U}(-3,3)$ & $0.79^{+0.84}_{-0.40}$ \\
    $m_1$ ($M_{\rm J}$) & mass & -- & $34^{+29}_{-11}$ \\
    $a_1$ (au) & semimajor axis & -- & $6.6^{+4.8}_{-2.4}$ \\
    $P_1$ (yr) & period & -- & $12^{+15}_{-6}$ \\
    $d_{1,\rm peri}$ (au) & pericenter distance & $\log\mathcal{U}(-2,2)$ & $3.7^{+3.7}_{-2.4}$ \\
    $\sqrt{e_1}\cos\omega_1$ & eccentricity-related component & $\mathcal{U}(-1,1)$ & $-0.2^{+0.5}_{-0.5}$\\
    $\sqrt{e_1}\sin\omega_1$ & eccentricity-related component & $\mathcal{U}(-1,1)$ & $-0.1^{+0.5}_{-0.5}$\\
    $e_1$ & eccentricity & -- & $0.43^{+0.34}_{-0.30}$ \\
    $\omega_1$ & argument of pericenter & -- & $3.4^{+1.5}_{-1.5}$ \\
    $\Omega_1$ & longitude of ascending node & $\mathcal{U}(0,2\pi)$ & $1.1^{+2.8}_{-0.8}$ \\
    $i_1$ & inclination & $\mathcal{U}(0,\pi)$ & $1.5^{+1.0}_{-0.9}$ \\
    $M_{1,0}$ & mean anomaly at reference time $t_0$ & $\mathcal{U}(0,2\pi)$ & $4.0^{+1.1}_{-2.6}$\\
    $\mu_{\alpha*,0}$ ($\rm \frac{mas}{yr}$) & systemic proper motion (RA)& $\mathcal{U}(1,5)$ & $3.251^{+0.030}_{-0.008}$ \\
    $\mu_{\delta,0}$ ($\rm \frac{mas}{yr}$) & systemic proper motion (Dec) & $\mathcal{U}(-7,-5)$ & $-6.01^{+0.03}_{-0.03}$\\
    \hline
    \end{tabular}
    \label{tab:prior_posterior}
\end{table*}

\begin{figure}
    \centering
    \includegraphics[width=1.0\linewidth]{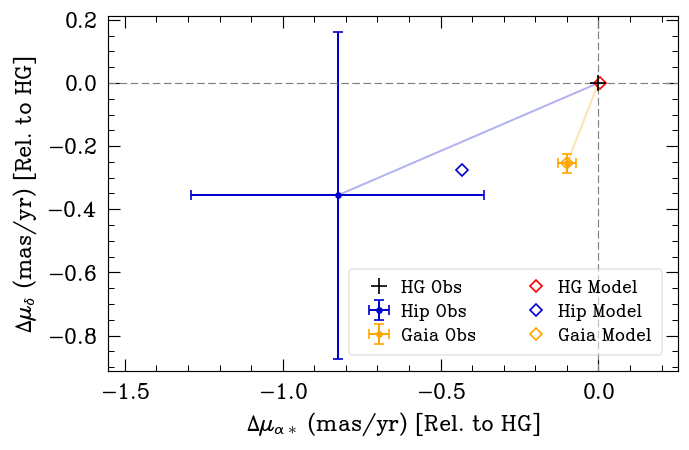}
    \caption{Error bars show the observed Hipparcos (blue) and Gaia (orange) proper motions relative to the HG proper motion (chosen to be the origin because
    it has the highest precision). Best-fit model predictions are shown as hollow diamonds.}
    \label{fig:HGCA_residuals}
\end{figure}

Figure~\ref{fig:HGCA_schematic} shows the stellar orbit in the best-fit solution, Figure~\ref{fig:TTV_residuals} presents the best-fit model for the TTV data, and Figure~\ref{fig:HGCA_residuals} compares the observed and modeled proper motions.
The best-fit model yields $\chi^2 = \chi^2_{\rm ast} + \chi^2_{\rm ttv} = 0.6 + 48.6 = 49.2$ for 38 degrees of freedom. Table~\ref{tab:prior_posterior} lists the posterior constraints for all model parameters and relevant derived quantities.
Figure~\ref{fig:corner} in Appendix~\ref{app:sup} presents the posterior distributions of selected orbital parameters of the companion from the combined runs, along with histograms from the individual runs.

The posterior favors a companion with a mass
in the brown-dwarf regime and a semimajor axis $\sim 7$~au. 
Although a tail of solutions extends
to higher companion masses and wider orbits,
high-contrast imaging non-detections with NIRC2 have tentatively ruled out companions with masses $\gtrsim100~M_{\rm J}$ at projected separations $\gtrsim100$~au \citep{Zhang_2026}, disfavoring
companion masses in the stellar regime and wider
orbits.

Although several orbital parameters -- particularly the eccentricity -- are weakly constrained, certain derived combinations are well determined.
The astrometric signal directly probes
the host star's acceleration, which scales as $m_1/a_1^2$. We
find $m_1/a_1^2 = 0.79^{+0.84}_{-0.40}~M_{\rm J}/{\rm au}^{2}$. 
The TTV data provide complementary constraints. Because the observations span a pericenter passage (2023--2024; see Figure~\ref{fig:TTV_residuals}), the timing variations are especially sensitive to the companion's pericenter distance. We infer $d_{1,\mathrm{peri}} = 3.7_{-2.4}^{+3.7}$~au.

\section{Formation Implications}

\subsection{Dynamical instability beyond the ice line}
\label{sec:stability}

With a luminosity of $L_\star \approx 15~L_\odot$ \citep{Stassun_2019},
the ice line of KELT-20 is expected at orbital distances $\sim$8--15~au \citep[e.g.,][]{Ida_2005,Kennedy_2008,vant_Hoff_2022,Kim_2025}, as compared to $\sim$3~au around Sun-like stars.
If KELT-20\,b formed beyond the ice line, its orbit would likely have crossed that of the companion, whose pericenter distance is of order 4~au.

A related question is how far from the star the proto--hot Jupiter
could have resided while remaining dynamically stable in the presence
of the companion.
\citet{Holman_1999} derived an empirical criterion for the long-term stability of a circular, circumstellar, coplanar planet in a binary system.  
The maximum stable planetary semimajor axis, based on integrations over $10^4$ binary orbits, is given by
\begin{equation}
\begin{aligned}
    \frac{a_b}{a_1} \lesssim ~ &0.464 - 0.380\mu - 0.631e_1 + 0.586\mu e_1 \\
    &+ 0.150e_1^2 - 0.198\mu e_1^2.\label{eq:HW99}
\end{aligned}
\end{equation}

The age of KELT-20 is $58\pm 5$~Myr, much longer than $10^4$ binary orbits for a companion with an orbital period of 
order 10 years. If the orbits are far from coplanar, dynamical
stability lifetimes are generally shorter \citep[e.g.][]{Quarles_2020}. 
Equation~\eqref{eq:HW99} therefore
provides a reasonable (and likely conservative) upper bound on the maximum stable planetary semimajor axis over the system's lifetime. 
Evaluating this criterion across
the posterior distribution of companion orbits
(Figure~\ref{fig:corner}), we find $a_{b,\max} = 1.5^{+1.6}_{-1.1}$~au (1$\sigma$), with $a_{b,\max} < 6.9$~au at the 3$\sigma$ level.

Even at the upper end of this range, the region
of long-term stability lies well interior to the
expected location of the ice line. This indicates that any long-lived planet in the system
must have either formed interior to the ice line or migrated inward from larger distances
in concert with the companion. In the
following sections, we examine both possibilities and find that formation interior to the ice line is more consistent with the available constraints.

\subsection{Formation inside the ice line} \label{ssec:formation_inside}

Assuming the companion reached its present-day orbit prior to the formation of KELT-20\,b, its presence would strongly inhibit planet formation beyond the ice line.
Within the standard core-accretion framework,
this poses a challenge: the surface density of solids interior to the ice line is typically too low to assemble a $\sim$10~$M_{\oplus}$ core rapidly enough to trigger runaway gas accretion before the gas disk dissipates \citep{pollack1996accretion,ida2004deterministic,armitage2010pf}.

Pebble-driven core accretion models might provide a way around this difficulty. \citet{Johnston_2024} identified a formation channel that operates
around massive stars but not around solar-mass hosts. In their simulations, embryos seeded at small orbital separations ($r_0 \lesssim 1$~au) begin with very low masses ($\lesssim 10^{-5}~M_\oplus$),
leading to slow initial growth and
delaying the onset of efficient inward migration until the gas has been substantially depleted. Because migration of the embryos
was found to be less efficient around massive stars -- due to the weaker torques resulting from a lower planet-to-star mass ratio, as well as shorter disk lifetimes -- the embryos can grow to $\gtrsim10^2~M_\oplus$ while avoiding rapid inward migration to the disk inner edge, where accretion would otherwise be halted. In contrast, around solar-mass stars, the simulated migration rates are faster, and embryos can reach the inner disk edge before attaining giant-planet masses.

This mechanism allows for giant planet
formation interior to the ice line
around massive stars.
However, the simulations of
\citet{Johnston_2024} did not produce
gas giants at orbital distances as small as $0.1$~au
before disk dispersal, indicating that
\textit{in situ} formation of hot Jupiters remains unlikely and that some subsequent migration mechanism is still required.

One possible pathway is that KELT-20\,b formed at $\sim1$~au via pebble-driven core accretion
and later underwent von Zeipel-Lidov-Kozai (ZLK) oscillations driven by the companion, leading to high-eccentricity migration. This scenario requires a large mutual inclination
($\gtrsim 60^\circ$ in this specific system; \citealt{Liu_2015}) and
typically produces spin-orbit misalignment
\citep[e.g.,][]{Fabrycky2007}. The observed sky-projected obliquity of KELT-20\,b is instead very small, making this pathway less likely. 

Alternatively, if the mutual inclination is small, the planet could undergo coplanar high-eccentricity migration, provided the companion's orbit
is sufficiently eccentric \citep{Petrovich_2015}. In this case, the planetary orbit could remain aligned with the stellar spin axis.

\subsection{Formation outside the ice line} \label{ssec:formation_outside}

The inference that KELT-20\,b formed interior to the ice line rests
on the assumption that the companion was already near its present-day orbit
at the time of planet formation. Here we briefly consider alternative scenarios in which the companion formed at larger distances and migrated inward later,
thereby avoiding early dynamical instability and permitting a more traditional formation pathway beyond the ice line.

One possibility is that the planet and additional massive companions
formed at larger orbital distances, and subsequent dynamical instabilities scattered both KELT-20\,b and its detected companion inward \citep[e.g.,][]{rice2003substellar}. In this picture,
KELT-20\,b would eventually reach its current orbit through tidal circularization \citep[e.g.,][]{Nagasawa2008,Beauge2012}. Such a scenario
likely requires the ejection of multiple massive bodies and would tend to produce spin–orbit misalignment \citep[e.g.,][]{Lu2025}. The small observed projected obliquity therefore argues against this possibility.

Another possibility is that the planet and the detected massive companion
underwent simultaneous disk-driven migration. If the proto--hot Jupiter formed near 10~au, the companion would need an initial semimajor axis of at least $\sim$20~au to remain dynamically stable with the planet. However, 
\citet{Armitage_2002} showed that brown-dwarf–mass companions at separations $\gtrsim 14$~au are unlikely to have had enough time
to migrate significantly before the disk disperses, making this scenario less likely.

\section{Discussion and Conclusions} \label{sec:disc}

The KELT-20 system consists of an A2 star,
a transiting ultra-hot Jupiter, and a massive companion newly detected from a joint
analysis of astrometric and transit-timing
data. Although the companion's parameters remain weakly constrained, its mass is probably in the brown-dwarf regime and its pericenter distance is $d_{1,\mathrm{peri}} = 3.7^{+3.7}_{-2.4}$~au (see Appendix~\ref{app:sup}, Figure~\ref{fig:corner}), although we cannot fully rule out a stellar-mass companion on a wider orbit. Forthcoming time-series astrometry from Gaia DR4 and DR5 should decisively distinguish between the brown-dwarf and stellar-mass families of solutions (see Appendix~\ref{app:sup}, Figure~\ref{fig:gaia_mock}). In recovery tests assuming the best-fit low-mass solution as the injected signal, the companion semimajor axis is recovered to within 10--40\% by DR4, depending on inclination, and to better than 1\% by DR5.

Radial velocities (RVs) are expected to provide very limited additional leverage. KELT-20 is a rapidly rotating A-type star, making precise Doppler measurements difficult due to few sharp absorption lines and substantial rotational broadening. 
The RVs reported by \citet{Lund_2017} span only $\sim$30 days, and have uncertainties of a few hundred m\,s$^{-1}$. The available RV data are therefore limited both by their precision and by their time baseline, and do not meaningfully constrain the long-period companion.

Based on the current observational constraints,
if the companion has remained near its present-day orbit since formation,
KELT-20\,b must have formed within ${\sim}3.7$~au to avoid orbit crossing and within ${\sim}1.5$~au to remain dynamically stable over the system's
lifetime. Under either criterion, formation beyond the water-ice line is disfavored, presenting a challenge for standard giant-planet formation scenarios.

These conclusions depend on the assumption that the companion has not undergone substantial orbital
evolution. Alternative scenarios remain viable in which both the planet and the companion formed at larger orbital distances and subsequently migrated inward, thereby avoiding early dynamical instability.
Atmospheric abundances may offer an independent probe of KELT-20\,b's formation
history. Existing measurements of volatile and refractory species \citep{Chachan_2025}, as well as planetary C/O ratios \citep{Fu_2022, Finnerty_2025} already provide useful constraints.
Improved measurements of the planetary C/O ratio, together with constraints
on the host star's elemental abundances, would help test whether KELT-20\,b
formed near the ice line or at smaller orbital distances.

KELT-20 brings together observational clues about hot Jupiter formation that are rarely available for a single system: astrometric acceleration, transit-timing variations, long-term dynamical constraints, and atmospheric
abundance measurements. If the companion has remained
near its present orbit since formation, the current data already preclude the possibility that KELT-20\,b 
formed beyond the water-ice line. With forthcoming Gaia astrometry
and prospects for improved atmospheric spectra, KELT-20
offers a promising opportunity
to test where hot Jupiters can --- and cannot --- be born.

\begin{acknowledgements}
    We thank Phil Armitage for fruitful discussions and invaluable feedback throughout the course of this work. We are also grateful to David Hogg, Megan Bedell, Jerry Xuan, Mordecai-Mark Mac Low, Ryan Rubenzahl, Quang Tran, Arjun Savel, Yubo Su, Chris Wang, Jiaxuan Li, and Caleb Lammers for their helpful comments. This work benefited from insightful conversations with the CCA Exoplanets \& Planet Formation Groups, the Winn Research Group, and the Rice Research Group. T.L.\ is supported by a Flatiron Research Fellowship at the Flatiron Institute, a division of the Simons Foundation. D.A.Y. is supported by a Juan Carlos Torres Postdoctoral Fellowship at the Massachusetts Institute of Technology.
\end{acknowledgements}

\software{\texttt{astropy} \citep{astropy}, \texttt{dynesty} \citep{dynesty}, \texttt{jaxoplanet} \citep{jaxoplanet}, \texttt{gaiascanlaw} \citep{gaiascanlaw_paper}, \texttt{smplotlib} \citep{smplotlib}}

\newpage
\appendix
\section{Supplementary Data and Figures}
\label{app:sup}

As noted in Section~\ref{ssec:TTV}, we exclude mid-transit times other than those obtained from TESS and HST from our joint analysis. For completeness, the previously reported measurements are listed in Table~\ref{tab:past_data}.

\begin{table}[h!]
    \caption{Mid-transit times of KELT-20\,b from the literature that are excluded from the joint astrometry-TTV analysis.}
    \centering
    \begin{tabular}{l r r l r}
    \hline
    Epoch & $T_{\rm mid}$ (BJD-2457000) & $\sigma$ (min) & Instrument & Reference \\
    \hline
    -611 & 301.621915 & 1.23 & WCO & \citet{Lund_2017} \\
    -541 & 544.810920 & 0.73 & DEMONEXT & \citet{Lund_2017} \\
    -444 & 881.795676 & 1.25 & WCO & \citet{Lund_2017} \\
    -444 & 881.796903 & 0.92 & CDK20N & \citet{Lund_2017} \\
    -444 & 881.799595 & 0.80 & PvdK & \citet{Lund_2017} \\
    -442 & 888.745551 & 0.85 & WCO & \citet{Lund_2017} \\
    -436 & 909.5907 & 0.0004 & IAC80 & \citet{Talens_2018} \\
    -434 & 916.537500 & 0.83 & CROW & \citet{Lund_2017} \\
    -320 & 1312.58566 & 0.00022 & MuSCAT2 & \citet{MuSCAT2_2019}\\
    -5 & 2406.927174 & 0.000024 & CHEOPS & \citet{CHEOPS_2024} \\
    4 & 2438.194141 & 0.000032 & CHEOPS & \citet{CHEOPS_2024} \\
    10 & 2459.038579 & 0.000251 & CHEOPS & \citet{CHEOPS_2024} \\
    96 & 2757.811176 & 0.000019 & PEPSI & \citet{PEPSI_2026}\\
    105 & 2789.078614 & 0.00044 & CHEOPS & \citet{CHEOPS_2024} \\
    \hline
    \end{tabular}
    \label{tab:past_data}
\end{table}

\begin{figure*} 
    \centering
    \includegraphics[width=0.9\linewidth]{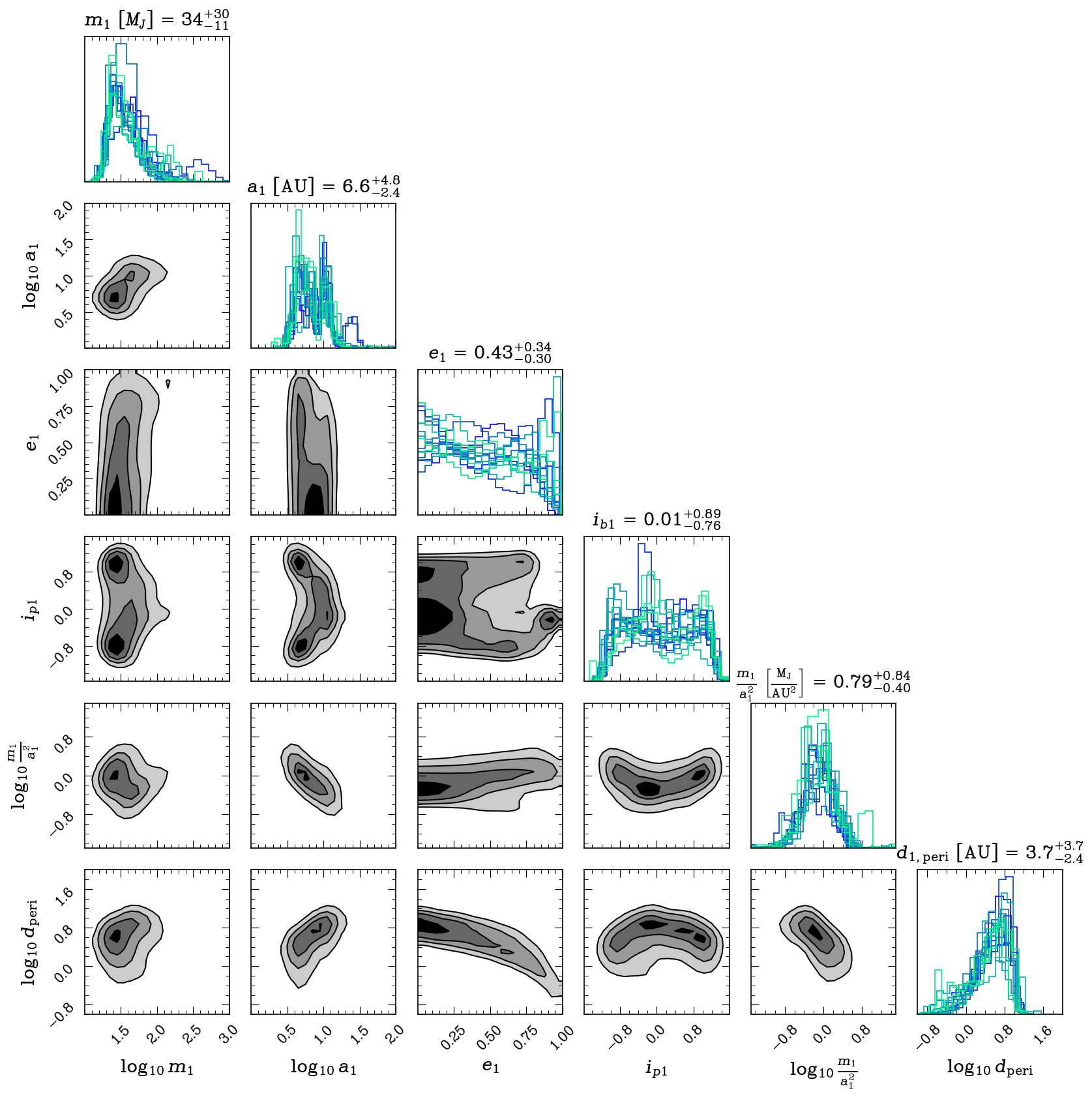}
    \caption{Posterior distributions of the companion's mass, semimajor axis, eccentricity, mutual inclination relative to the transiting planet, astrometric acceleration measure, and pericenter distance, as inferred from the joint astrometric and TTV analysis. Filled contours indicate 0.5, 1, 1.5, and 2-$\sigma$ confidence levels. Colored histograms correspond to independent sampling runs. Values quoted above each diagonal panel give the posterior median and central 68\% credible interval, defined by the 16th and 84th percentiles of the combined posterior.}
    \label{fig:corner}
\end{figure*}

Figure~\ref{fig:corner} shows the posterior distributions of selected orbital parameters of the companion
along with histograms from the individual runs,
based on the joint analysis in Section~\ref{ssec:joint}. 

\begin{figure}
    \centering
    \includegraphics[width=0.6\columnwidth]{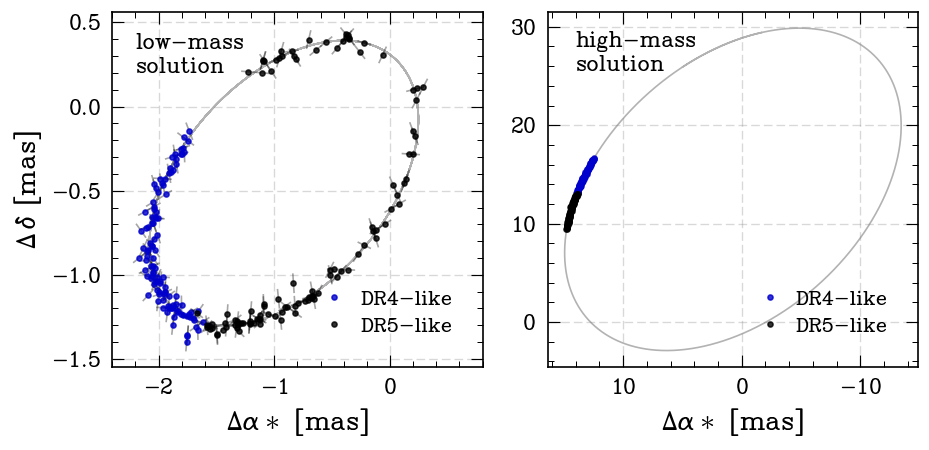}
    \caption{Simulated Gaia astrometric data of the KELT-20 host star for best-fit brown-dwarf and stellar-companion solutions. 
    Blue and black points are mock DR4 and DR5 measurements. Gray curves indicate the underlying model reflex motion. The origin is the center of mass.
    \textbf{Left:} Expected DR4 and DR5 orbital coverage of the host star under the perturbation of a brown-dwarf companion ($m_1 = 56~\rm M_J$, $a_1 = 6.3$~au, $e_1=0.82$). 
    \textbf{Right:} Expected orbital coverage in the presence of a stellar companion ($m_1 = 0.17~\rm M_\odot$, $a_1 = 11$~au, $e_1=0.94$).}
    \label{fig:gaia_mock}
\end{figure}

\newpage
We simulated Gaia-like astrometric time series by generating Keplerian orbits with \texttt{jaxoplanet}\footnote{\url{https://github.com/exoplanet-dev/jaxoplanet}} \citep{jaxoplanet} and propagating posterior orbital solutions to the DR4 (5.5 year baseline) and DR5 (10 year baseline) epochs, adopting the Gaia scanning law implemented in \texttt{gaiascanlaw} \citep{gaiascanlaw_paper} and converting along-scan measurements to sky-plane positions following \citet{Yahalomi2026_resoeccentric}. Figure~\ref{fig:gaia_mock} shows the predicted astrometric signals for the best-fit brown-dwarf and stellar-companion solutions as discussed in Section~\ref{sec:disc}.

\newpage 

\bibliography{Bib}

@ARTICLE{TESS,
       author = {{Ricker}, George R. and {Winn}, Joshua N. and {Vanderspek}, Roland and {Latham}, David W. and {Bakos}, G{\'a}sp{\'a}r {\'A}. and {Bean}, Jacob L. and {Berta-Thompson}, Zachory K. and {Brown}, Timothy M. and {Buchhave}, Lars and {Butler}, Nathaniel R. and {Butler}, R. Paul and {Chaplin}, William J. and {Charbonneau}, David and {Christensen-Dalsgaard}, J{\o}rgen and {Clampin}, Mark and {Deming}, Drake and {Doty}, John and {De Lee}, Nathan and {Dressing}, Courtney and {Dunham}, Edward W. and {Endl}, Michael and {Fressin}, Francois and {Ge}, Jian and {Henning}, Thomas and {Holman}, Matthew J. and {Howard}, Andrew W. and {Ida}, Shigeru and {Jenkins}, Jon M. and {Jernigan}, Garrett and {Johnson}, John Asher and {Kaltenegger}, Lisa and {Kawai}, Nobuyuki and {Kjeldsen}, Hans and {Laughlin}, Gregory and {Levine}, Alan M. and {Lin}, Douglas and {Lissauer}, Jack J. and {MacQueen}, Phillip and {Marcy}, Geoffrey and {McCullough}, Peter R. and {Morton}, Timothy D. and {Narita}, Norio and {Paegert}, Martin and {Palle}, Enric and {Pepe}, Francesco and {Pepper}, Joshua and {Quirrenbach}, Andreas and {Rinehart}, Stephen A. and {Sasselov}, Dimitar and {Sato}, Bun'ei and {Seager}, Sara and {Sozzetti}, Alessandro and {Stassun}, Keivan G. and {Sullivan}, Peter and {Szentgyorgyi}, Andrew and {Torres}, Guillermo and {Udry}, Stephane and {Villasenor}, Joel},
        title = "{Transiting Exoplanet Survey Satellite (TESS)}",
      journal = {Journal of Astronomical Telescopes, Instruments, and Systems},
         year = 2015,
        month = jan,
       volume = {1},
          eid = {014003},
        pages = {014003},
          doi = {10.1117/1.JATIS.1.1.014003},
       adsurl = {https://ui.adsabs.harvard.edu/abs/2015JATIS...1a4003R}
}

@ARTICLE{Kennedy_2008,
       author = {{Kennedy}, Grant M. and {Kenyon}, Scott J.},
        title = "{Planet Formation around Stars of Various Masses: The Snow Line and the Frequency of Giant Planets}",
      journal = {\apj},
         year = 2008,
        month = jan,
       volume = {673},
       number = {1},
        pages = {502-512},
          doi = {10.1086/524130},
archivePrefix = {arXiv},
       eprint = {0710.1065},
 primaryClass = {astro-ph},
       adsurl = {https://ui.adsabs.harvard.edu/abs/2008ApJ...673..502K}
}

@article{vant_Hoff_2022,
doi = {10.3847/1538-4357/ac3080},
url = {https://doi.org/10.3847/1538-4357/ac3080},
year = {2022},
month = {jan},
publisher = {The American Astronomical Society},
volume = {924},
number = {1},
pages = {5},
author = {van ’t Hoff, Merel L. R. and Harsono, Daniel and van Gelder, Martijn L. and Hsieh, Tien-Hao and Tobin, John J. and Jensen, Sigurd S. and Hirano, Naomi and Jørgensen, Jes K. and Bergin, Edwin A. and van Dishoeck, Ewine F.},
title = {Imaging the Water Snowline around Protostars with Water and HCO+ Isotopologues},
journal = {The Astrophysical Journal}
}

@ARTICLE{Distler_2026,
       author = {{Distler}, Adam and {Soares-Furtado}, Melinda and {Mann}, Andrew W. and {Kraus}, Adam L. and {Gagn{\'e}}, Jonathan and {Becker}, Juliette and {Narayan}, Ritvik Sai and {Clark}, Max and {Vanderburg}, Andrew and {Rodriguez}, Joseph E. and {Rogers}, Laura K. and {Kerr}, Ronan},
        title = "{TESS Hunt for Young and Maturing Exoplanets (THYME). XIV. A Comoving-based Age Constraint for KELT-20}",
      journal = {\aj},
         year = 2026,
        month = apr,
       volume = {171},
       number = {4},
          eid = {248},
        pages = {248},
          doi = {10.3847/1538-3881/ae4c48},
archivePrefix = {arXiv},
       eprint = {2603.01313},
 primaryClass = {astro-ph.EP},
       adsurl = {https://ui.adsabs.harvard.edu/abs/2026AJ....171..248D}
}

@ARTICLE{Kim_2025,
       author = {{Kim}, Young-Jun and {Lee}, Jeong-Eun and {Baek}, Giseon and {Lee}, Seokho},
        title = "{Water Snowline in Young Stellar Objects with Various Density Structures Using Radiative Transfer Models}",
      journal = {Journal of Korean Astronomical Society},
         year = 2025,
        month = oct,
       volume = {58},
        pages = {243-254},
          doi = {10.5303/JKAS.2025.58.2.243},
archivePrefix = {arXiv},
       eprint = {2510.14294},
 primaryClass = {astro-ph.SR},
       adsurl = {https://ui.adsabs.harvard.edu/abs/2025JKAS...58..243K}
}

@ARTICLE{Ida_2005,
       author = {{Ida}, S. and {Lin}, D.~N.~C.},
        title = "{Toward a Deterministic Model of Planetary Formation. III. Mass Distribution of Short-Period Planets around Stars of Various Masses}",
      journal = {\apj},
         year = 2005,
        month = jun,
       volume = {626},
       number = {2},
        pages = {1045-1060},
          doi = {10.1086/429953},
archivePrefix = {arXiv},
       eprint = {astro-ph/0502566},
 primaryClass = {astro-ph},
       adsurl = {https://ui.adsabs.harvard.edu/abs/2005ApJ...626.1045I}
}

@ARTICLE{Stassun_2019,
       author = {{Stassun}, Keivan G. and {Oelkers}, Ryan J. and {Paegert}, Martin and {Torres}, Guillermo and {Pepper}, Joshua and {De Lee}, Nathan and {Collins}, Kevin and {Latham}, David W. and {Muirhead}, Philip S. and {Chittidi}, Jay and {Rojas-Ayala}, B{\'a}rbara and {Fleming}, Scott W. and {Rose}, Mark E. and {Tenenbaum}, Peter and {Ting}, Eric B. and {Kane}, Stephen R. and {Barclay}, Thomas and {Bean}, Jacob L. and {Brassuer}, C.~E. and {Charbonneau}, David and {Ge}, Jian and {Lissauer}, Jack J. and {Mann}, Andrew W. and {McLean}, Brian and {Mullally}, Susan and {Narita}, Norio and {Plavchan}, Peter and {Ricker}, George R. and {Sasselov}, Dimitar and {Seager}, S. and {Sharma}, Sanjib and {Shiao}, Bernie and {Sozzetti}, Alessandro and {Stello}, Dennis and {Vanderspek}, Roland and {Wallace}, Geoff and {Winn}, Joshua N.},
        title = "{The Revised TESS Input Catalog and Candidate Target List}",
      journal = {\aj},
         year = 2019,
        month = oct,
       volume = {158},
       number = {4},
          eid = {138},
        pages = {138},
          doi = {10.3847/1538-3881/ab3467},
archivePrefix = {arXiv},
       eprint = {1905.10694},
 primaryClass = {astro-ph.SR},
       adsurl = {https://ui.adsabs.harvard.edu/abs/2019AJ....158..138S}
}

@ARTICLE{Johnston_2024,
       author = {{Johnston}, Heather F. and {Pani{\'c}}, O. and {Liu}, B.},
        title = "{Formation of giant planets around intermediate-mass stars}",
      journal = {\mnras},
         year = 2024,
        month = jan,
       volume = {527},
       number = {2},
        pages = {2303-2322},
          doi = {10.1093/mnras/stad3254},
archivePrefix = {arXiv},
       eprint = {2310.17767},
 primaryClass = {astro-ph.EP},
       adsurl = {https://ui.adsabs.harvard.edu/abs/2024MNRAS.527.2303J}
}

@article{Quarles_2020,
doi = {10.3847/1538-3881/ab64fa},
url = {https://doi.org/10.3847/1538-3881/ab64fa},
year = {2020},
month = {feb},
publisher = {The American Astronomical Society},
volume = {159},
number = {3},
pages = {80},
author = {Quarles, Billy and Li, Gongjie and Kostov, Veselin and Haghighipour, Nader},
title = {Orbital Stability of Circumstellar Planets in Binary Systems},
journal = {The Astronomical Journal}
}

@ARTICLE{Fu_2022,
       author = {{Fu}, Guangwei and {Sing}, David K. and {Lothringer}, Joshua D. and {Deming}, Drake and {Ih}, Jegug and {Kempton}, Eliza M.-R. and {Malik}, Matej and {Komacek}, Thaddeus D. and {Mansfield}, Megan and {Bean}, Jacob L.},
        title = "{Strong H$_{2}$O and CO Emission Features in the Spectrum of KELT-20b Driven by Stellar UV Irradiation}",
      journal = {\apjl},
         year = 2022,
        month = jan,
       volume = {925},
       number = {1},
          eid = {L3},
        pages = {L3},
          doi = {10.3847/2041-8213/ac4968},
archivePrefix = {arXiv},
       eprint = {2201.02261},
 primaryClass = {astro-ph.EP},
       adsurl = {https://ui.adsabs.harvard.edu/abs/2022ApJ...925L...3F}
}

@article{Liu_2015,
       author = {{Liu}, Bin and {Mu{\~n}oz}, Diego J. and {Lai}, Dong},
        title = "{Suppression of extreme orbital evolution in triple systems with short-range forces}",
      journal = {\mnras},
         year = 2015,
        month = feb,
       volume = {447},
       number = {1},
        pages = {747-764},
          doi = {10.1093/mnras/stu2396},
archivePrefix = {arXiv},
       eprint = {1409.6717},
 primaryClass = {astro-ph.EP},
       adsurl = {https://ui.adsabs.harvard.edu/abs/2015MNRAS.447..747L}
}

@article{Mandel2002,
doi = {10.1086/345520},
url = {https://dx.doi.org/10.1086/345520},
year = {2002},
month = {nov},
publisher = {},
volume = {580},
number = {2},
pages = {L171},
author = {Mandel, Kaisey and Agol, Eric},
title = {Analytic Light Curves for Planetary Transit Searches},
journal = {The Astrophysical Journal}
}

@ARTICLE{Petrovich_2015,
       author = {{Petrovich}, Cristobal},
        title = "{Hot Jupiters from Coplanar High-eccentricity Migration}",
      journal = {\apj},
         year = 2015,
        month = may,
       volume = {805},
       number = {1},
          eid = {75},
        pages = {75},
          doi = {10.1088/0004-637X/805/1/75},
archivePrefix = {arXiv},
       eprint = {1409.8296},
 primaryClass = {astro-ph.EP},
       adsurl = {https://ui.adsabs.harvard.edu/abs/2015ApJ...805...75P}
}

@article{Fabrycky2007,
doi = {10.1086/521702},
url = {https://dx.doi.org/10.1086/521702},
year = {2007},
month = {nov},
publisher = {},
volume = {669},
number = {2},
pages = {1298},
author = {Fabrycky, Daniel and Tremaine, Scott},
title = {Shrinking Binary and Planetary Orbits by Kozai Cycles with Tidal Friction*},
journal = {The Astrophysical Journal}
}

@article{Wu2003,
doi = {10.1086/374598},
url = {https://dx.doi.org/10.1086/374598},
year = {2003},
month = {may},
publisher = {},
volume = {589},
number = {1},
pages = {605},
author = {Wu, Y. and Murray, N.},
title = {Planet Migration and Binary Companions: The Case of HD 80606b},
journal = {The Astrophysical Journal}
}

@ARTICLE{Armitage_2002,
       author = {{Armitage}, Philip J. and {Bonnell}, Ian A.},
        title = "{The brown dwarf desert as a consequence of orbital migration}",
      journal = {\mnras},
         year = 2002,
        month = feb,
       volume = {330},
       number = {1},
        pages = {L11-L14},
          doi = {10.1046/j.1365-8711.2002.05213.x},
archivePrefix = {arXiv},
       eprint = {astro-ph/0112001},
 primaryClass = {astro-ph},
       adsurl = {https://ui.adsabs.harvard.edu/abs/2002MNRAS.330L..11A}
}

@ARTICLE{Finnerty_2025,
       author = {{Finnerty}, Luke and {Xin}, Yinzi and {Xuan}, Jerry W. and {Inglis}, Julie and {Fitzgerald}, Michael P. and {Agrawal}, Shubh and {Baker}, Ashley and {Bartos}, Randall and {Blake}, Geoffrey A. and {Calvin}, Benjamin and {Cetre}, Sylvain and {Delorme}, Jacques-Robert and {Doppmann}, Greg and {Echeverri}, Daniel and {Horstman}, Katelyn and {Hsu}, Chih-Chun and {Jovanovic}, Nemanja and {Liberman}, Joshua and {L{\'o}pez}, Ronald A. and {Martin}, Emily C. and {Mawet}, Dimitri and {Morris}, Evan and {Pezzato}, Jacklyn and {Ruffio}, Jean-Baptiste and {Sappey}, Ben and {Schofield}, Tobias and {Skemer}, Andrew and {Venenciano}, Taylor and {Wallace}, J. Kent and {Wallack}, Nicole L. and {Wang}, Jason J. and {Wang}, Ji},
        title = "{Water Dissociation and Rotational Broadening in the Atmosphere of KELT-20 b from High-resolution Spectroscopy}",
      journal = {\aj},
         year = 2025,
        month = jun,
       volume = {169},
       number = {6},
          eid = {333},
        pages = {333},
          doi = {10.3847/1538-3881/adce02},
archivePrefix = {arXiv},
       eprint = {2503.01946},
 primaryClass = {astro-ph.EP},
       adsurl = {https://ui.adsabs.harvard.edu/abs/2025AJ....169..333F}
}

@ARTICLE{Lund_2017,
       author = {{Lund}, Michael B. and {Rodriguez}, Joseph E. and {Zhou}, George and {Gaudi}, B. Scott and {Stassun}, Keivan G. and {Johnson}, Marshall C. and {Bieryla}, Allyson and {Oelkers}, Ryan J. and {Stevens}, Daniel J. and {Collins}, Karen A. and {Penev}, Kaloyan and {Quinn}, Samuel N. and {Latham}, David W. and {Villanueva}, Jr., Steven and {Eastman}, Jason D. and {Kielkopf}, John F. and {Oberst}, Thomas E. and {Jensen}, Eric L.~N. and {Cohen}, David H. and {Joner}, Michael D. and {Stephens}, Denise C. and {Relles}, Howard and {Corfini}, Giorgio and {Gregorio}, Joao and {Zambelli}, Roberto and {Esquerdo}, Gilbert A. and {Calkins}, Michael L. and {Berlind}, Perry and {Ciardi}, David R. and {Dressing}, Courtney and {Patel}, Rahul and {Gagnon}, Patrick and {Gonzales}, Erica and {Beatty}, Thomas G. and {Siverd}, Robert J. and {Labadie-Bartz}, Jonathan and {Kuhn}, Rudolf B. and {Col{\'o}n}, Knicole D. and {James}, David and {Pepper}, Joshua and {Fulton}, Benjamin J. and {McLeod}, Kim K. and {Stockdale}, Christopher and {Calchi Novati}, Sebastiano and {DePoy}, D.~L. and {Gould}, Andrew and {Marshall}, Jennifer L. and {Trueblood}, Mark and {Trueblood}, Patricia and {Johnson}, John A. and {Wright}, Jason and {McCrady}, Nate and {Wittenmyer}, Robert A. and {Johnson}, Samson A. and {Sergi}, Anthony and {Wilson}, Maurice and {Sliski}, David H.},
        title = "{KELT-20b: A Giant Planet with a Period of P {\ensuremath{\sim}} 3.5 days Transiting the V {\ensuremath{\sim}} 7.6 Early A Star HD 185603}",
      journal = {\aj},
         year = 2017,
        month = nov,
       volume = {154},
       number = {5},
          eid = {194},
        pages = {194},
          doi = {10.3847/1538-3881/aa8f95},
archivePrefix = {arXiv},
       eprint = {1707.01518},
 primaryClass = {astro-ph.EP},
       adsurl = {https://ui.adsabs.harvard.edu/abs/2017AJ....154..194L}
}

@ARTICLE{Batygin_2016,
       author = {{Batygin}, Konstantin and {Bodenheimer}, Peter H. and {Laughlin}, Gregory P.},
        title = "{In Situ Formation and Dynamical Evolution of Hot Jupiter Systems}",
      journal = {\apj},
         year = 2016,
        month = oct,
       volume = {829},
       number = {2},
          eid = {114},
        pages = {114},
          doi = {10.3847/0004-637X/829/2/114},
archivePrefix = {arXiv},
       eprint = {1511.09157},
 primaryClass = {astro-ph.EP},
       adsurl = {https://ui.adsabs.harvard.edu/abs/2016ApJ...829..114B}
}

@ARTICLE{Lee_2016,
       author = {{Lee}, Eve J. and {Chiang}, Eugene},
        title = "{Breeding Super-Earths and Birthing Super-puffs in Transitional Disks}",
      journal = {\apj},
         year = 2016,
        month = feb,
       volume = {817},
       number = {2},
          eid = {90},
        pages = {90},
          doi = {10.3847/0004-637X/817/2/90},
archivePrefix = {arXiv},
       eprint = {1510.08855},
 primaryClass = {astro-ph.EP},
       adsurl = {https://ui.adsabs.harvard.edu/abs/2016ApJ...817...90L}
}

@ARTICLE{Boley_2016,
       author = {{Boley}, A.~C. and {Granados Contreras}, A.~P. and {Gladman}, B.},
        title = "{The In Situ Formation of Giant Planets at Short Orbital Periods}",
      journal = {\apjl},
         year = 2016,
        month = feb,
       volume = {817},
       number = {2},
          eid = {L17},
        pages = {L17},
          doi = {10.3847/2041-8205/817/2/L17},
archivePrefix = {arXiv},
       eprint = {1510.04276},
 primaryClass = {astro-ph.EP},
       adsurl = {https://ui.adsabs.harvard.edu/abs/2016ApJ...817L..17B}
}

@ARTICLE{Ida_2008,
       author = {{Ida}, S. and {Lin}, D.~N.~C.},
        title = "{Toward a Deterministic Model of Planetary Formation. IV. Effects of Type I Migration}",
      journal = {\apj},
         year = 2008,
        month = jan,
       volume = {673},
       number = {1},
        pages = {487-501},
          doi = {10.1086/523754},
archivePrefix = {arXiv},
       eprint = {0802.1114},
 primaryClass = {astro-ph},
       adsurl = {https://ui.adsabs.harvard.edu/abs/2008ApJ...673..487I}
}

@ARTICLE{Heller_2019,
       author = {{Heller}, Ren{\'e}},
        title = "{Formation of hot Jupiters through disk migration and evolving stellar tides}",
      journal = {\aap},
         year = 2019,
        month = aug,
       volume = {628},
          eid = {A42},
        pages = {A42},
          doi = {10.1051/0004-6361/201833486},
archivePrefix = {arXiv},
       eprint = {1806.06601},
 primaryClass = {astro-ph.EP},
       adsurl = {https://ui.adsabs.harvard.edu/abs/2019A&A...628A..42H}
}

@INPROCEEDINGS{Baruteau_2014,
       author = {{Baruteau}, C. and {Crida}, A. and {Paardekooper}, S.-J. and {Masset}, F. and {Guilet}, J. and {Bitsch}, B. and {Nelson}, R. and {Kley}, W. and {Papaloizou}, J.},
        title = "{Planet-Disk Interactions and Early Evolution of Planetary Systems}",
    booktitle = {Protostars and Planets VI},
         year = 2014,
       editor = {{Beuther}, Henrik and {Klessen}, Ralf S. and {Dullemond}, Cornelis P. and {Henning}, Thomas},
        month = jan,
        pages = {667-689},
          doi = {10.2458/azu_uapress_9780816531240-ch029},
archivePrefix = {arXiv},
       eprint = {1312.4293},
 primaryClass = {astro-ph.EP},
       adsurl = {https://ui.adsabs.harvard.edu/abs/2014prpl.conf..667B}
}

@ARTICLE{Wu2011,
       author = {{Wu}, Yanqin and {Lithwick}, Yoram},
        title = "{Secular Chaos and the Production of Hot Jupiters}",
      journal = {\apj},
         year = 2011,
        month = jul,
       volume = {735},
       number = {2},
          eid = {109},
        pages = {109},
          doi = {10.1088/0004-637X/735/2/109},
archivePrefix = {arXiv},
       eprint = {1012.3475},
 primaryClass = {astro-ph.EP},
       adsurl = {https://ui.adsabs.harvard.edu/abs/2011ApJ...735..109W}
}

@ARTICLE{Beauge2012,
       author = {{Beaug{\'e}}, C. and {Nesvorn{\'y}}, D.},
        title = "{Multiple-planet Scattering and the Origin of Hot Jupiters}",
      journal = {\apj},
         year = 2012,
        month = jun,
       volume = {751},
       number = {2},
          eid = {119},
        pages = {119},
          doi = {10.1088/0004-637X/751/2/119},
archivePrefix = {arXiv},
       eprint = {1110.4392},
 primaryClass = {astro-ph.EP},
       adsurl = {https://ui.adsabs.harvard.edu/abs/2012ApJ...751..119B}
}

@ARTICLE{Nagasawa2008,
       author = {{Nagasawa}, M. and {Ida}, S. and {Bessho}, T.},
        title = "{Formation of Hot Planets by a Combination of Planet Scattering, Tidal Circularization, and the Kozai Mechanism}",
      journal = {\apj},
         year = 2008,
        month = may,
       volume = {678},
       number = {1},
        pages = {498-508},
          doi = {10.1086/529369},
archivePrefix = {arXiv},
       eprint = {0801.1368},
 primaryClass = {astro-ph},
       adsurl = {https://ui.adsabs.harvard.edu/abs/2008ApJ...678..498N}
}

@ARTICLE{Rafikov_2005,
       author = {{Rafikov}, Roman R.},
        title = "{Can Giant Planets Form by Direct Gravitational Instability?}",
      journal = {\apjl},
         year = 2005,
        month = mar,
       volume = {621},
       number = {1},
        pages = {L69-L72},
          doi = {10.1086/428899},
archivePrefix = {arXiv},
       eprint = {astro-ph/0406469},
 primaryClass = {astro-ph},
       adsurl = {https://ui.adsabs.harvard.edu/abs/2005ApJ...621L..69R}
}

@ARTICLE{MuSCAT2_2019,
       author = {{Casasayas-Barris}, N. and {Pall{\'e}}, E. and {Yan}, F. and {Chen}, G. and {Kohl}, S. and {Stangret}, M. and {Parviainen}, H. and {Helling}, Ch. and {Watanabe}, N. and {Czesla}, S. and {Fukui}, A. and {Monta{\~n}{\'e}s-Rodr{\'\i}guez}, P. and {Nagel}, E. and {Narita}, N. and {Nortmann}, L. and {Nowak}, G. and {Schmitt}, J.~H.~M.~M. and {Zapatero Osorio}, M.~R.},
        title = "{Atmospheric characterization of the ultra-hot Jupiter MASCARA-2b/KELT-20b. Detection of CaII, FeII, NaI, and the Balmer series of H (H{\ensuremath{\alpha}}, H{\ensuremath{\beta}}, and H{\ensuremath{\gamma}}) with high-dispersion transit spectroscopy}",
      journal = {\aap},
         year = 2019,
        month = aug,
       volume = {628},
          eid = {A9},
        pages = {A9},
          doi = {10.1051/0004-6361/201935623},
archivePrefix = {arXiv},
       eprint = {1905.12491},
 primaryClass = {astro-ph.EP},
       adsurl = {https://ui.adsabs.harvard.edu/abs/2019A&A...628A...9C}
}

@ARTICLE{PEPSI_2026,
       author = {{Lenhart}, Calder and {Johnson}, Marshall C. and {Wang}, Ji and {Asnodkar}, Anusha Pai and {Petz}, Sydney and {Duck}, Alison and {Strassmeier}, Klaus G. and {Ilyin}, Ilya},
        title = "{PEPSI Investigation, Retrieval, and Atlas of Numerous Giant Atmospheres (PIRANGA). II. Phase-resolved Cross-correlation Transmission Spectroscopy of KELT-20b}",
      journal = {\aj},
         year = 2026,
        month = feb,
       volume = {171},
       number = {2},
          eid = {81},
        pages = {81},
          doi = {10.3847/1538-3881/ae2252},
archivePrefix = {arXiv},
       eprint = {2503.07719},
 primaryClass = {astro-ph.EP},
       adsurl = {https://ui.adsabs.harvard.edu/abs/2026AJ....171...81L}
}

@ARTICLE{CHEOPS_2024,
       author = {{Singh}, V. and {Scandariato}, G. and {Smith}, A.~M.~S. and {Cubillos}, P.~E. and {Lendl}, M. and {Billot}, N. and {Fortier}, A. and {Queloz}, D. and {Sousa}, S.~G. and {Csizmadia}, Sz. and {Brandeker}, A. and {Carone}, L. and {Wilson}, T.~G. and {Akinsanmi}, B. and {Patel}, J.~A. and {Krenn}, A. and {Demangeon}, O.~D.~S. and {Bruno}, G. and {Pagano}, I. and {Hooton}, M.~J. and {Cabrera}, J. and {Santos}, N.~C. and {Alibert}, Y. and {Alonso}, R. and {Asquier}, J. and {B{\'a}rczy}, T. and {Navascues}, D. Barrado and {Barros}, S.~C.~C. and {Baumjohann}, W. and {Beck}, M. and {Beck}, T. and {Benz}, W. and {Bergomi}, M. and {Bonfanti}, A. and {Bonfils}, X. and {Borsato}, L. and {Broeg}, C. and {Charnoz}, S. and {Cameron}, A. Collier and {Davies}, M.~B. and {Deleuil}, M. and {Deline}, A. and {Delrez}, L. and {Demory}, B.-O. and {Ehrenreich}, D. and {Erikson}, A. and {Fossati}, L. and {Fridlund}, M. and {Gandolfi}, D. and {Gillon}, M. and {G{\"u}del}, M. and {G{\"u}nther}, M.~N. and {Harre}, J.-V. and {Heitzmann}, A. and {Helling}, Ch. and {Hoyer}, S. and {Isaak}, K.~G. and {Kiss}, L.~L. and {Lam}, K.~W.~F. and {Laskar}, J. and {des Etangs}, A. Lecavelier and {Magrin}, D. and {Maxted}, P.~F.~L. and {Mischler}, H. and {Mordasini}, C. and {Nascimbeni}, V. and {Olofsson}, G. and {Ottensamer}, R. and {Pall{\'e}}, E. and {Peter}, G. and {Piotto}, G. and {Pollacco}, D. and {Ragazzoni}, R. and {Rando}, N. and {Rauer}, H. and {Ribas}, I. and {Salmon}, S. and {S{\'e}gransan}, D. and {Simon}, A.~E. and {Stalport}, M. and {Steinberger}, M. and {Szab{\'o}}, Gy. M. and {Thomas}, N. and {Udry}, S. and {Ulmer}, B. and {Van Grootel}, V. and {Venturini}, J. and {Villaver}, E. and {Walton}, N.~A. and {Zingales}, T.},
        title = "{CHEOPS observations of KELT-20 b/MASCARA-2 b: An aligned orbit and signs of variability from a reflective day side}",
      journal = {\aap},
         year = 2024,
        month = mar,
       volume = {683},
          eid = {A1},
        pages = {A1},
          doi = {10.1051/0004-6361/202347533},
archivePrefix = {arXiv},
       eprint = {2311.03264},
 primaryClass = {astro-ph.EP},
       adsurl = {https://ui.adsabs.harvard.edu/abs/2024A&A...683A...1S}
}

@ARTICLE{Toomre_1964,
       author = {{Toomre}, A.},
        title = "{On the gravitational stability of a disk of stars.}",
      journal = {\apj},
         year = 1964,
        month = may,
       volume = {139},
        pages = {1217-1238},
          doi = {10.1086/147861},
       adsurl = {https://ui.adsabs.harvard.edu/abs/1964ApJ...139.1217T}
}

@ARTICLE{Perri_1974,
       author = {{Perri}, F. and {Cameron}, A.~G.~W.},
        title = "{Hydrodynamic Instability of the Solar Nebula in the Presence of a Planetary Core}",
      journal = {\icarus},
         year = 1974,
        month = aug,
       volume = {22},
       number = {4},
        pages = {416-425},
          doi = {10.1016/0019-1035(74)90074-8},
       adsurl = {https://ui.adsabs.harvard.edu/abs/1974Icar...22..416P}
}

@article{Lu2025,
doi = {10.3847/1538-4357/ad9b79},
url = {https://dx.doi.org/10.3847/1538-4357/ad9b79},
year = {2025},
month = {jan},
publisher = {The American Astronomical Society},
volume = {979},
number = {2},
pages = {218},
author = {Lu, Tiger and An, Qier and Li, Gongjie and Millholland, Sarah C. and Rice, Malena and Brandt, G. Mirek and Brandt, Timothy D.},
title = {Planet–Planet Scattering and Von Zeipel–Lidov–Kozai Migration—The Dynamical History of HAT-P-11},
journal = {The Astrophysical Journal}
}

@ARTICLE{astropy,
       author = {{Astropy Collaboration} and {Price-Whelan}, Adrian M. and {Lim}, Pey
       Lian and {Earl}, Nicholas and {Starkman}, Nathaniel and {Bradley}, Larry and
       {Shupe}, David L. and {Patil}, Aarya A. and {Corrales}, Lia and {Brasseur}, C.~E.
       and {N{\"o}the}, Maximilian and {Donath}, Axel and {Tollerud}, Erik and {Morris},
       Brett M. and {Ginsburg}, Adam and {Vaher}, Eero and {Weaver}, Benjamin A. and
       {Tocknell}, James and {Jamieson}, William and {van Kerkwijk}, Marten H. and
       {Robitaille}, Thomas P. and {Merry}, Bruce and {Bachetti}, Matteo and
       {G{\"u}nther}, H. Moritz and {Aldcroft}, Thomas L. and {Alvarado-Montes}, Jaime
       A. and {Archibald}, Anne M. and {B{\'o}di}, Attila and {Bapat}, Shreyas and
       {Barentsen}, Geert and {Baz{\'a}n}, Juanjo and {Biswas}, Manish and {Boquien},
       M{\'e}d{\'e}ric and {Burke}, D.~J. and {Cara}, Daria and {Cara}, Mihai and
       {Conroy}, Kyle E. and {Conseil}, Simon and {Craig}, Matthew W. and {Cross},
       Robert M. and {Cruz}, Kelle L. and {D'Eugenio}, Francesco and {Dencheva}, Nadia
       and {Devillepoix}, Hadrien A.~R. and {Dietrich}, J{\"o}rg P. and {Eigenbrot},
       Arthur Davis and {Erben}, Thomas and {Ferreira}, Leonardo and {Foreman-Mackey},
       Daniel and {Fox}, Ryan and {Freij}, Nabil and {Garg}, Suyog and {Geda}, Robel and
       {Glattly}, Lauren and {Gondhalekar}, Yash and {Gordon}, Karl D. and {Grant},
       David and {Greenfield}, Perry and {Groener}, Austen M. and {Guest}, Steve and
       {Gurovich}, Sebastian and {Handberg}, Rasmus and {Hart}, Akeem and
       {Hatfield-Dodds}, Zac and {Homeier}, Derek and {Hosseinzadeh}, Griffin and
       {Jenness}, Tim and {Jones}, Craig K. and {Joseph}, Prajwel and {Kalmbach}, J.
       Bryce and {Karamehmetoglu}, Emir and {Ka{\l}uszy{\'n}ski}, Miko{\l}aj and
       {Kelley}, Michael S.~P. and {Kern}, Nicholas and {Kerzendorf}, Wolfgang E. and
       {Koch}, Eric W. and {Kulumani}, Shankar and {Lee}, Antony and {Ly}, Chun and
       {Ma}, Zhiyuan and {MacBride}, Conor and {Maljaars}, Jakob M. and {Muna}, Demitri
       and {Murphy}, N.~A. and {Norman}, Henrik and {O'Steen}, Richard and {Oman}, Kyle
       A. and {Pacifici}, Camilla and {Pascual}, Sergio and {Pascual-Granado}, J. and
       {Patil}, Rohit R. and {Perren}, Gabriel I. and {Pickering}, Timothy E. and
       {Rastogi}, Tanuj and {Roulston}, Benjamin R. and {Ryan}, Daniel F. and {Rykoff},
       Eli S. and {Sabater}, Jose and {Sakurikar}, Parikshit and {Salgado}, Jes{\'u}s
       and {Sanghi}, Aniket and {Saunders}, Nicholas and {Savchenko}, Volodymyr and
       {Schwardt}, Ludwig and {Seifert-Eckert}, Michael and {Shih}, Albert Y. and
       {Jain}, Anany Shrey and {Shukla}, Gyanendra and {Sick}, Jonathan and {Simpson},
       Chris and {Singanamalla}, Sudheesh and {Singer}, Leo P. and {Singhal}, Jaladh and
       {Sinha}, Manodeep and {Sip{\H{o}}cz}, Brigitta M. and {Spitler}, Lee R. and
       {Stansby}, David and {Streicher}, Ole and {{\v{S}}umak}, Jani and {Swinbank},
       John D. and {Taranu}, Dan S. and {Tewary}, Nikita and {Tremblay}, Grant R. and
       {Val-Borro}, Miguel de and {Van Kooten}, Samuel J. and {Vasovi{\'c}}, Zlatan and
       {Verma}, Shresth and {de Miranda Cardoso}, Jos{\'e} Vin{\'\i}cius and {Williams},
       Peter K.~G. and {Wilson}, Tom J. and {Winkel}, Benjamin and {Wood-Vasey}, W.~M.
       and {Xue}, Rui and {Yoachim}, Peter and {Zhang}, Chen and {Zonca}, Andrea and
       {Astropy Project Contributors}}, title = "{The Astropy Project: Sustaining and
       Growing a Community-oriented Open-source Project and the Latest Major Release
       (v5.0) of the Core Package}",
      journal = {\apj},
         year = 2022,
        month = aug,
       volume = {935},
       number = {2},
          eid = {167},
        pages = {167},
          doi = {10.3847/1538-4357/ac7c74},
archivePrefix = {arXiv},
       eprint = {2206.14220},
 primaryClass = {astro-ph.IM},
       adsurl = {https://ui.adsabs.harvard.edu/abs/2022ApJ...935..167A}
}

@software{jaxoplanet,
  author       = {Soichiro Hattori and
                  Lionel Garcia and
                  Catriona Murray and
                  Jiayin Dong and
                  Shashank Dholakia and
                  David Degen and
                  Daniel Foreman-Mackey},
  title        = {{exoplanet-dev/jaxoplanet: Astronomical time series analysis with JAX}},
  month        = mar,
  year         = 2024,
  publisher    = {Zenodo},
  version      = {v0.0.2},
  doi          = {10.5281/zenodo.10736936},
  url          = {https://doi.org/10.5281/zenodo.10736936}
}

@software{smplotlib,
  author       = {Jiaxuan Li},
  title        = {AstroJacobLi/smplotlib: v0.0.9},
  month        = jul,
  year         = 2023,
  publisher    = {Zenodo},
  version      = {v0.0.9},
  doi          = {10.5281/zenodo.8126529},
  url          = {https://doi.org/10.5281/zenodo.8126529},
}

@article{Bodenheimer_2000,
title = {Models of the in Situ Formation of Detected Extrasolar Giant Planets,},
journal = {Icarus},
volume = {143},
number = {1},
pages = {2-14},
year = {2000},
issn = {0019-1035},
doi = {https://doi.org/10.1006/icar.1999.6246},
url = {https://www.sciencedirect.com/science/article/pii/S0019103599962462},
author = {Peter Bodenheimer and Olenka Hubickyj and Jack J. Lissauer}}

@ARTICLE{Ivshina2022,
       author = {{Ivshina}, Ekaterina S. and {Winn}, Joshua N.},
        title = "{TESS Transit Timing of Hundreds of Hot Jupiters}",
      journal = {\apjs},
         year = 2022,
        month = apr,
       volume = {259},
       number = {2},
          eid = {62},
        pages = {62},
          doi = {10.3847/1538-4365/ac545b},
archivePrefix = {arXiv},
       eprint = {2202.03401},
 primaryClass = {astro-ph.EP},
       adsurl = {https://ui.adsabs.harvard.edu/abs/2022ApJS..259...62I}
}

@article{Brandt_2021,
doi = {10.3847/1538-4365/abf93c},
url = {https://doi.org/10.3847/1538-4365/abf93c},
year = {2021},
month = {jun},
publisher = {The American Astronomical Society},
volume = {254},
number = {2},
pages = {42},
author = {Brandt, Timothy D.},
title = {The Hipparcos–Gaia Catalog of Accelerations: Gaia EDR3 Edition},
journal = {The Astrophysical Journal Supplement Series}
}

@ARTICLE{dynesty,
       author = {{Speagle}, Joshua S.},
        title = "{DYNESTY: a dynamic nested sampling package for estimating Bayesian posteriors and evidences}",
      journal = {\mnras},
         year = 2020,
        month = apr,
       volume = {493},
       number = {3},
        pages = {3132-3158},
          doi = {10.1093/mnras/staa278},
archivePrefix = {arXiv},
       eprint = {1904.02180},
 primaryClass = {astro-ph.IM},
       adsurl = {https://ui.adsabs.harvard.edu/abs/2020MNRAS.493.3132S}
}

@ARTICLE{Higson_2019,
       author = {{Higson}, Edward and {Handley}, Will and {Hobson}, Mike and {Lasenby}, Anthony},
        title = "{Dynamic nested sampling: an improved algorithm for parameter estimation and evidence calculation}",
      journal = {Statistics and Computing},
         year = 2019,
        month = sep,
       volume = {29},
       number = {5},
        pages = {891-913},
          doi = {10.1007/s11222-018-9844-0},
archivePrefix = {arXiv},
       eprint = {1704.03459},
 primaryClass = {stat.CO},
       adsurl = {https://ui.adsabs.harvard.edu/abs/2019S&C....29..891H}
}
\bibliographystyle{aasjournal}

\end{CJK*}
\end{document}